\documentclass[12pt]{article}

\usepackage{amsmath}
\usepackage{amssymb}
\usepackage{amsthm}
\usepackage{amscd}
\usepackage{amsfonts}
\usepackage{dsfont}
\usepackage{cases}
\usepackage[applemac]{inputenc}
\usepackage{enumerate}
\usepackage[all]{xy}
\usepackage{esint}
\usepackage{graphicx}
\usepackage{url}
\theoremstyle{plain}

\theoremstyle{definition}

\newcommand{\N}{\mathbb N}

\newcommand{\E}{\mathrm E}

\begin{document}

\title{A Mathematical Approach to Comply with Ethical \\Constraints in Compassionate Use Treatments}
\date{}
\author{F. Thomas Bruss \\Universit\'e Libre de Bruxelles}
\maketitle

\begin{abstract}
\noindent Patients who are seriously ill may ask doctors to treat them with unapproved medication, about which not much is known, or else with known medication in a  high dosage. 
Apart from strict legal constraints such cases may  involve difficult ethical questions as e.g.\,how long a series of treatments of different patients  should be continued. Similar questions also arise in less serious situations. A physician trusts that a certain combination of freely available drugs are efficient against a specific disease and tries to help patients and to follow at the same time the {\it primum-non-nocere} principle. 

The objective of this paper is to contribute to the research on such questions in the form of mathematical models. Arguing in a step-to-step approach, we will show that certain sequential optimisation problems comply  in a natural way with the true spirit of major ethical principles in medicine. We then suggest protocols and associate algorithms to find optimal, or approximately optimal, treatment strategies. Although the contribution may sometimes be difficult to apply in medical practice, the author thinks that the rational behind the approach offers a valuable alternative for finding decision support and should attract attention.
\medskip


\end{abstract}

\section{Introduction}
It is not evident that, apart from statistical conclusions, Mathematics may play an important role for questions of ethical behaviour in medical practice. Nevertheless, there are  situations which lead to a type of difficult ethical questions where Mathematics can help. We speak in particular about questions which arise in {\it compassionate use} medical treatments, also called {\it expanded access} treatments. 

Patients who are seriously ill may come to see doctors, often in an act of last hope, to treat them with unapproved medication about which not much is known, or else previously known medication, however in a unapproved high dosage. In these cases one speaks of requests of a {compassionate use} or, synonymously, of {expanded access} treatments.
\subsection{Legal aspects}
In most countries specific legal constraints apply for requests of compassionate use treatments. The common denominator behind the legal constraints is the intention to protect patients. This intention usually translates, often word by word, into the foremost ethical principle of medicine that the expected benefit of a treatment should outweigh the risk for a patient, or somewhat extended, outweigh the patient's loss of quality of life. 

 One cannot expect to find in different legislations  a one-to-one correspondence of sub-points of the intention to observe the ethical principle. In fact, even in societies which are expected to be similar in their traditional and cultural background, its interpretation and thus its impact on legal constraints can vary considerably. 
See e.g.  Darrow et al (2015), mainly concentrating on priorities in the US and relevant FDA-regulations, and  Sou (2010), summarising EU compassionate use programmes.
In most countries, a written declaration of patients is required, sometimes even with the additional requirement of being certified by a third party. 

\subsection{Ethical aspects}
The intention that the expected benefit of a treatment of a patient should outweigh the risk for this patient  is in the same way a substantial part in the codes of medical conduct and medical professionalism. This is evident from the ancient oath of Hippocrates (460 BC - 370 BC) over the writings of Percival (1740-1804) up to new codes such as the {\it Declaration of a new doctor}; (see e.g. Sritharan et al. (2001).)

MacKenzie (2007) gives a clear and organised overview of the subject of medical professionalism, dealing with several of these questions. Expanded access and compassionate use  programs for experimental drugs 
and their characterisation are also well presented  in the article by Miller et al. (2017). \,One notices also that compassionate use of investigational drugs seems to attract increased interest in recent times; Bunnik et al. (2017) relate this with a broader a change in the {\it landscape} of un-met medical needs. We think in this paper mainly of drugs; for related questions, centered around alternative procedures in surgery, see e.g. Angelos (2010). 

If the physician agrees to a compassionate use program, then, legal aspects left aside, ethical questions quickly take a specific meaning. How long should the treatment with possible heavy side effects be continued? In a sequence of patients awaiting the treatment, should the doctor decide to stop (no further treatments) if no success has been observed so far among the patients? Similar questions for a conscientious doctor also arise in less serious situations. For instance, the physician would like to test the efficiency of combinations of approved treatments with known side effects. The {\it primum-non-nocere} principle, a pillar in ethical medical codes, implies similar questions in a natural way, and
patients would typically rely on the doctor to take the decisions. 

\subsection {Searching for guidelines}
The conscientious doctor (he, say) may have no more than these traditional guidelines, however. As far as compassionate use is concerned, he typically does not know enough about the efficiency of the drug. He must try to learn from observations of treated patients controlled with untreated patients. If there are several requests of a similar type, he can hope to learn  from sequential observation. 
But then, what objective function should be optimised, and what guidelines can the doctor hope for?

These questions, shortly addressed in Bruss (2006),  can be put in mathematical models which we  present below. They are kept natural and modest in the hypotheses so that they may be used under relatively general conditions. Moreover, for these models the answers can be found in a rather straightforward manner by the {\it odds-algorithm} and its modifications. In principle they can always be applied, because independence is the main requirement to apply it, and different patients react typically independently of each other.  

\smallskip As far as the author is aware, neither such models  nor the methodology based on the odds-algorithm has attracted interest in the medical community. This can in part be explained
by the fact that compassionate use treatments represent fortunately only a small percentage of medical treatments. Still, physicians confronted with such problems may strongly appreciate advice.

 The goal of the present article is  to draw interest of the medical community to this method, and this now for two more reasons. First, the odds algorithm has in the meantime been applied in different domains, such of search strategies, investment strategies, and robotic maintenance problems (see e.g. Ano et al. (2010), Tamaki (2010)
 and  Dendievel (2013) for an overview.) Second, 
 several improvements have been found in different directions which make both the models and the algorithms more tractable for the real-world medical practice.
 
\subsection{From guidelines to a first model}
The idea is to develop a suitable model in a step-to-step approach.

Suppose there are ten, say, patients who ask for a compassionate use treatment. To begin, suppose for simplicity, that these patients are all in a comparable state of health and that the outcome of a treatment can be classified in a reasonable time span as being  a {\it success} (= clear improvement of the state of health) or else a {\it failure} (= no improvement or deterioration.) If the doctor treated all patients sequentially he would then see at the end a sequence of ten {\bf+} or  {\bf -}. He cannot predict what will happen. For learning what may happen the following Gedankenspiel may be motivating, however:

\subsection{Prophetical abilities and probabilistic reasoning}

{\bf (i) The prophet.}  Suppose for a moment that the doctor was a prophet,
 able to foresee the future of outcomes. Suppose he saw that, if he treated all ten,  there would be two, say, successes among them. Suppose moreover that as a true prophet he knows the names of the fortunate patients. Then he treats of course only these two patients, and informs  the other eight that they should not undergo the treatment in question. For such a prophet the situation is trivial; he has no decision problem.
  
\bigskip \noindent
{\bf(ii) The half-prophet.}
Let us weaken the preceding hypothesis. Our physician is now supposed to be only a {\it half-prophet} in the sense that he would know that there will be two successes but neither know the names of the two fortunate ones  nor their position in the queue. He could then announce the chance for success 1 in 5 to the patients and see who would like to stay to participate under this condition. 
Suppose nine of them would stay, and that after five treatments the sequence of outcomes were
\begin{align} -,  +,  -,  -,  + , ~ ?_6 ,  \,?_7 , \,?_8 , \,?_9 \end{align}
Here the conscientious physician would stop treatments with the second success, i.e. with the fifth patient, knowing that the following four would suffer in vain from a useless treatment.  
Indeed, note that the last "+" in such an imaginary sequence, wherever it may be located, plays  
a  distinguished role: \begin{quote} The {\it last} success in a sequence of treatments is the {\it first} one  to {\it complete} the subset of {\it all}  treatments resulting in a success.\end{quote}
 It is true that if the two success had appeared earlier the doctor could have spared useless treatments to more patients. However, being only a half-prophet the doctor could not know. From the ethical point of view his behaviour is irreproachable.

\bigskip \noindent
{\bf (iii) The physician.}
In reality, the doctor is neither a prophet nor a half-prophet. However, we see here an important part of the ethical directive which follows from the logic of the Gedankenspiel. Indeed, as long as patients stay in the sequence,  the doctor should try to {\it sequentially maximise  the probability} of recognising the last + in the sequence of treatments. The reasons are twofold, namely
\begin{quote}
\medskip(a) the last + is the first treatment covering {\it all successful treatments} in the given sequence.
Sparing treatments for the following patients
increases their quality of life and renders them free to
envisage alternatives.

\medskip
(b) the probability of a further + is {\it strictly decreasing} with each further treatment.\end{quote}

\medskip
\noindent Statement (a) is evident. (b) is not trivial, and the proof given in Bruss (2000), (see p.\,1386). 
Interestingly, (b) is true
even if the doctor has no idea about the success probabilities for the different patients. The essence of the proof lies in the following.
 
\smallskip
\noindent{\bf Theorem }:  Suppose that there are $n$ patients
for whom the success probabilities {\it were known}
to be $p_1, p_2, \cdots, p_n.$ Then, with $q_k=1-p_k,~ r_k=p_k/q_k$ and $1\le s \le n,$ 

\begin{quote}
(i)  The function $V(n,s)$ defined by \begin{align}V(n,s):=\prod_{k=s}^n q_k \sum_{j=1}^n r_j\end{align} represents the probability that the first + from patient number $s$ onward is the very last + in the sequence of the $n$ patients.

(ii) The function $V(n,s)$ is {\it unimodal} in 
$s\in\{1,2, \cdots, n\}.$
\end{quote}

\noindent Unimodal means here that $V(n,s)$ either decreases for all $s,$ or else it increases up to some $s_{\rm max}$ and then decreases thereafter. If the probabilities $p_k$ and thus the odds  $r_k$ are known then (Bruss (2000)) the index $s$ maximizing $V(n,s)$
is determined by
\begin{align}s = \begin{cases} &1, ~~\mbox{if }\sum_{j=1}^n r_j\le 1\\
&\mbox{largest $k$ such that} \sum_{j=k}^n r_j\ge 1
,~~\mbox{otherwise.}\end{cases}\end{align}
Thus with known success probabilities, the first ethical directive for the physician is clear. He should treat sequentially the patients up to number $s-1$ and then be prepared to stop with the next successful treatment, if any, from number $s$ onwards.

A second ethical directive may be needed in terms of a {\it lower threshold} probability which tells when to stop if the  probability of a further success becomes too small. This lower threshold is fixed beforehand in agreement with the patients. Experience seems to indicate that patients in compassionate use treatments are often willing to accept particularly great risks
and tend to choose this threshold quite small.

We should also mention here that prior information can sometimes be used to improve the approach. Although patients are seen as individuals and thus independent of each other, the doctor may want to allow for dependence within groups of patients, for example according to sex, or according to having undergone the same history of other treatments. Ferguson (2016) studied the use the odds-algorithm also in the dependent case, allowing to take such situations into consideration. In the present paper, however, our interest is confined to the case of independent reactions of patients.\section{Protocols for specific situations}
In the following we shall show that {\it organised
protocols}, i.e.  step-to-step instructions and recommendations can, according to different situations, help to comply with both the patients' priorities and ethical constraints for the physician. This method is based on the odds algorithm (Bruss 2000) and refinements of this algorithm (Bruss (2005), Bruss and Louchard (2009), Dendievel (2012), and Ferguson (2016). The name odds-algorithm was coined by Bruss according to the important role played by the {\it odds} of events to find the optimal stopping time. The odds in the preceding Theorem 1 are the values $r_k=p_k/q_k.$ The odds of an event $E$ is simply the ratio $P(E)/(1-P(E))$ where $P(E)$ denotes the probability of an event $E.$ 
\subsection{Known success probabilities}
In the first protocol  we suppose that both the number of patients $n$ as well as their respective success probabilities are known. This is in practice not very realistic but it is the right introduction to show the essence of the method and the mathematical simplicity of the odds-algorithm. 

\subsection*{Protocol 1: ~Fixed number of patients;  success probability supposed to be known for each patient.} 

We set 

~~~~~$n =$ number of patients scheduled

~~~~~$p_k=$ success probability for the $k$th patient, $k=1, 2, \cdots, n.$

~~~~~$q_k=1-p_k, k=1, 2, \cdots, n.$

~~~~~$r_k=p_k/q_k, k=1, 2, \cdots, n.$
\subsection*{Algorithm for Protocol 1}

{\bf Step 1:} Write down {\it in reversed order} the $p_k, ~q_k $ and $r_k$ so that the line entries $q_k$ and $r_k$ are exactly under the line of the $p_k$, that is$$\rm{(i)}~~~~p_n,~~p_{n-1},~~p_{n-2}, ~~p_{n-3},~~ \cdots$$$$\rm{(ii)}~~~q_n,~~q_{n-1},~~q_{n-2}, ~~q_{n-3},~~ \cdots$$$$\rm{(iii)}~~~r_n,~~r_{n-1},~~\,r_{n-2},~~\,\, r_{n-3},~ \cdots.$$\smallskip
As defined above,  the entries of line (ii) are $1$ minus the entries of (i), and those of line (iii) the quotients of corresponding entries in (i) and (ii).

\medskip
\noindent{\bf Step 2:} Form sequentially the sums $r_n+r_{n-1}+r_{n-2}+\cdots$ and stop at that index $s$
where this sum reaches or exceeds for the {\it first} time the value 1. Otherwise, if the consecutive sums are too small to reach $1,$ then put $s=1.$  {\it The optimal policy is to stop on the first  success from $s$ onwards (if any).} We also obtain directly the corresponding optimal success probability. Hence with \begin{align}R(n,s)=r_n+r_{n-1}+\cdots+r_s\end{align}  either $R(n,s)\ge1$ for some intermediate $s,$ or else $R(n,1)<1$ and $s=1.$ Take note of this index $s$ and of $R(n,s)$ and compute from line (ii) the product
\begin{align} Q(n,s)=q_n \times q_{n-1} \times \cdots \times q_s. \end{align} Finally, multiplying (4) and (5) yields the probability of successfully  stopping on the very last $+$, that is \begin{align} V(n,s)=Q(n,s)R(n,s).~~ \qed\end{align}

This is all what has to be done to compute the optimal stopping index $s,$ and $V(n,s)$ is the probability of an overall success, i.e. having succeeded in obtaining all possible successes without any futile treatment thereafter. We give one numerical example.

\smallskip\noindent{\bf Example:} Suppose $n=7$  patients with success probabilities $p_1=.35, p_2=.1,  p_3=.05,  p_4=.3, p_5=.1, p_6=.15, p_7=.25$ queue for treatment. Then (i), (ii) and (iii) read in reversed order (and rounded to two decimals)
 $$~.25,~~.15,~~.1,~~.3,~~.05,~~.1,~~.35$$
 $$~.75,~~.85,~~.9,~~.7,~~.95,~~.9,~~.65$$
$$\,.33,~~.17,~\,.11,~.43, ~\cdots~ \cdots~ \cdots$$ 
According to (3) and (4), since $.33 +.17 + .11 =.61< 1$ but   $.33 +.17 + .11+.43 = 1.04 >1$, we obtain $s=4.$
Hence the doctor should treat patients numbers 1, 2, and 3
 in any case but then stop treatment after the next success, if any. The probability of stopping with the very last success (see (6)) is thus $ ~.75\times.85\times.9\times.7\times(1.04)\approx 0.418.$ 

\bigskip\noindent {\bf Remark}  ~One may wonder why in Step 1
the algorithm is given in terms of the reversed order of odds. An equivalent definition of the index $s$
in the natural order of the $r_1, r_2, \cdots,r_n$ is
\begin{align} s= \hbox{smallest index}~ k~ \hbox{such that} \sum_{j=k+1}^n r_j <1,~ 0 \le k \le n.\end{align} Indeed, our first formulation in (3) has the advantage that we only have to compute the odds as far as needed whereas, using (7), we first have to compute the sum of all odds and then withdraw successively $r_1, r_2, \cdots$ until the remaining sum drops below the value $1$. However, as  seen in the next protocol,  the definition (7) is sometimes more useful.

\subsubsection{Influence of the order of treatments}
We see from (i), (ii), (iii) and the definitions of $R(n,s)$ and $Q(n,s)$  in (4) and (5) that $V(n,s)$ may depend on the order of treatments, and in general it does. In our example, if the physician exchanges the places of the first and the fourth patient in the first line the values $.3$ and $.35$
are interchanged (recall the reversed order) so that now $p_4=.35,q_4=.65$ and thus $r_4\approx .55$
Hence $V(7,4)\approx (.75\times.85\times.9\times.65)\times(.33+.17+.11+.55)=0.428,$ thus a bit larger.
 
 \smallskip
There is no simple method to see  the best order because $V(n,s),$ although easy to compute  for a given order by this algorithm, is a relatively complicated discrete value function. Note that in theory there are $7!=5040$ possible different  permutations of treatments. However, in practice, time and other constraints on both sides,  doctor and patients, usually imply that there are only few possibilities to permute the order of treatments. For a few possible exchanges the values  $V(n,s)$ are then easy to check.
\subsubsection{Performance of the algorithm}
The algorithm is optimal, that is, one cannot do better
under the given hypotheses. Moreover, with the number of operations being sub-linear in $n$, there exist no quicker way to find $s$ and $V(n,s)$ at the same time, as proved in Bruss (2000).

\smallskip
We also know (Bruss (2003)) that, if the total sum of odds is at least $1$ then the optimal success probability is always above $1/e \approx 36.78\%.$ 
This may sound somewhat disappointing, but this is just the lower bound. In fact, the {\it effective} success rate is  in practice typically more, and this for two reasons. 
First, if $n$ is not large, $V(n,s)$ is usually above $40\%.$ Second, and in particular, the largest part of the theoretical loss of some $60\%$ lies in the non-existence of successes after index $s$ until the end. But it suffices for the doctor to look at $V(n,s+1), V(n,s+2), \cdots$ and to stop the sequence if these values drop below a lower threshold fixed in agreement with the patients. Stopping may miss further successes, of course, but sparing the following patients from the side-effects of increasingly likely useless treatments becomes an ethical priority.

\smallskip
We do not introduce a lower threshold  in Protocol 1. This is because 
the success probabilities are supposed to be known. The patients are informed and are of course always free to leave the queue, but future events are in this model independent of preceding observations.

\smallskip
A lower threshold for the probability of a further success may play an important role, if the success probabilities are not supposed to be known, as studied in the next Section. Protocol 1 is the simplest protocol incorporating the idea behind the odds-algorithm. Observation-independent success probabilities are often less realistic than unknown success probabilities. \section{Unknown success probabilities}
As argued before, known success probabilities are not the typical case in compassionate use treatments. Even if preceding parallel studies for a specific drug do exist, only rough estimates of these can be hoped for since sample sizes are usually  small. The most difficult situation for the physician is the one where almost nothing is known about the success probability of the drug. 
This is why we suppose that nothing at all is known about success probabilities. 

If $n=1$ the treatment of this  patient is then a deliberate trial and error, and it makes no sense to envisage a scheme of learning. However, if $n>1$, the  conscientious physician will start thinking about what statistical inference from preceding failures and successes might possibly contribute. We need a sufficiently simple model.

\subsection{Model of compound success probability}
Our way to model the unknown success probability for a given patient is to see it as the result of two agents (as proposed, in a preliminary form,  in Bruss (2006)).
The first agent is an unknown parameter $p,$ say, where $0<p<1,$
which is interpreted as the {\it internal} success probability of the drug. The second one is the general state of health of the patient. For a fixed number $n>1$ the physician is supposed to  give the patients $1, 2, \cdots, n$ scores or grades for their state of health $h_1, h_2, \cdots, h_n,$ where the two extreme values $h=100\%$ respectively $h=0\%$ are seen as formal upper and lower bound for the state of health of any person considering such a treatment.

\smallskip\noindent
 Writing $1$ for $100\%$ and $0$ for $0\%$ we 
now put formally \begin{align} p_k:=h_k\, p, ~k=1, 2,\cdots, n,\end{align} so that now $p_k$ is seen as the success probability of the treatment for patient number $k.$ Note that $h_k$ itself is not a probability but a constant chosen by the physician reducing success probability $p$ for patient number $k.$ We see $p$ as an unknown parameter and thus put $p_k=h_kp.$ If we knew $p$ to be a random variable governed by the law $\cal L$, say, we would see $p_k$ as random variable satisfying
\begin{align}\E_{\cal L}\left(p_k\right)=\E_{\cal L}\left(h_k\,p\right)=h_k\,\E_{\cal L}\left(p\right),\end{align} where $\E_{\cal L}$ denotes the expectation with respect to the measure $\cal L.$ In this case we could do better by a stepwise Bayesian inference after sequential observations, but the knowledge of $\cal L$ can hardly be expected, and so we do not study this case.

There is  clearly some arbitrariness for the physician in his choice of the constants $h_k.$  However, a physician would typically be able to judge whether the state of health of a patient compared to another is  much better, or, just a bit better.
If so, the physician should not hesitate too much to choose the $h_k$ and may also rely on his feelings. It is well known that health state interpretation can be  biased by external information, as it is, interestingly, also the case  for the interpretation of findings obtained in funded medical  research (see e.g Kesselheim et al. (2012)).  Actually,  since $p$ is not supposed to be known, it may be well justified for the doctor to fix only $h_{\rm min}=\min\{h_k: 1 \le k \le n\}$ and $h_{\rm max}=\max\{h_k: 1 \le k \le n\},$ and to spread the remaining values equally over $[h_{\rm min}, h_{\rm max}]$ according to the perceived relative ranks of the states of health. 

\smallskip
Now we are ready for Protocol 2. In order to facilitate the understanding of the general idea of the new situation and of the common parts with  Protocol 1, we will  not  incorporate the lower threshold  in this new protocol, but will do so in Protocal 3.
\subsubsection*{Protocol 2: $n$ known,  success probabilities unknown.} 

\begin{quote} $n=$ number of patients scheduled for treatment,\\$p=$ parameter thought of as being the unknown {\it internal} success probability of the drug for a randomly chosen patient,\\$h_k=$ physician's estimate of the state of health of patient number $k$ with $ 0<h_k<1,~ k=1, 2, \cdots, n,$\\$S_k= $number of successes (number of "+") up to the $k$th treatment. \end{quote}We now have the formal definition \begin{align}p_k=h_k\,p, ~~q_k=1-p_k, ~~r_k=p_k/q_k, \end{align}but with the difference that the success and failure probabilities $p_k$ and $q_k$ as well as the odds $r_k$ must be estimated sequentially, that is from the preceding observations up to patient number $k-1.$ A first indication for suitable estimators comes from the formal expectational argument, applied to  (8) and (9), namely
$$\E[S_k]=h_1\,p+h_2 \,p+ \cdots +h_k\,p=p(h_1+h_2+\cdots+h_k).$$ Hence, normalising at step $k$ the number of successes by $\sum_{j=1}^kh_j,$ we have
\begin{align}\E\left[\frac{S_k} {\sum_{j=1}^k h_j}\right] =p\end{align} so that ${S_k}/ {\sum_{j=1}^k h_j}$ is an {\it unbiased estimator} for the unknown $p $ at step $k$. It is moreover the {\it maximum likelihood estimator} of $p.$ Therefore we propose to use
at step $k$  \begin{align}\hat p_k:=\frac{S_k} {\sum_{j=1}^k h_j}\end{align} as the estimators for $p$ at step $k=1, 2, \cdots, n-1.$

\medskip \subsection{Future odds} Another impact of not knowing the success probabilities, and thus not knowing the odds, is that the index $s$ defined in (3) makes no longer sense. We use (7) because the odds must now be estimated  sequentially. If we replace the odds now by there estimates at step $k$ denoted by $ \hat r_j(k)$ say, we obtain correspondingly the instruction:
{Stop with the}~k\hbox{th treatment if}\begin{align}\sum_{j=k+1}^n \hat r_j(k) <1, ~ 1 \le k \le n.\end{align}

How would we estimate at stage $k$ these future odds $\hat r_j$ for successes at stages $j=k+1, k+2, \cdots, n-1$ given the history $S_1, S_2,\cdots , S_k$? Note that, at stage $k,$ we do not know $S_j$ for $j>k.$

We argue as follows. If $p$ were the intrinsic success probability of the treatment then the success probability at stage $j$ would equal $p_j=ph_j$ and the corresponding odds would be $r_j=ph_j/(1-ph_j).$ The last update on $p$ given $S_1, S_2,\cdots , S_k$ is given by (12) so that we propose to estimate the odds $r_j$ by  \begin{align}\hat r_j :=\frac{h_j \hat p_k}{1-h_j\hat p_k}.\end{align}
Putting $H_k:=h_1 + h_2+ \cdots +h_k,$ (12) becomes $\hat p_k= S_k/H_k$ so that, according to (14) the stopping index  (7) translates  into the rule :

\bigskip\noindent
\centerline{Stop the sequence with the $k$th treatment, $ 1 \le k \le n$, if}
\begin{align}
\sum_{j=k+1}^n \frac{h_jS_k}{\left[\left(H_k - h_j S_k\right)\right]^+} <1, ~ 1 \le k \le n,\end{align} where $[X]^+ = \max\{0,X\}. ~~\qed$ 

\medskip
Forgetting for the moment the function $[\,\cdot\,]^+,$ (15) becomes perfectly intuitive if we think for a moment that all $h_j$'s are equal. Each sum term in (15) then simplifies to $S_k/(k-S_k).$  This is always non-negative, since $k\ge S_k,$ so that $[\,\cdot\,]^+$ can be omitted. This is the "empirical odds" situation at step $k,$  that  is (7) where the $r_j$ are all replaced by the empirical odds at time $k.$ In the more typical cases of non-identical  $h_j$'s the $\hat r_j$'s are health-state weighted odds. 

To explain the truncation $[\,\cdot\,]^+$ note that, since the $S_k$'s are random variables, the denominators of the sum terms can  become negative for a relatively large $h_j$ if $S_k$ happens to be large. This can get in collision with the definition of odds. Thus, to stay consistent, we impose the truncation  $[\,\cdot\,]^+$  for all cases.

\smallskip
We also mention that the $\hat r_j$'s are slightly biased estimators of the $r_j$'s, but we use them  for simplicity. This seems justified since, in any case, the $h_j$'s 
bear some subjective bias from the physician's assessment of the health states of the patients.

\subsubsection{Beginning run of failures}
The condition in (14) is trivially satisfied as long as $S_k=0.$ This lies in the nature of the problem. As long as there are no successes there is, from a statistical point of view, no incentive to continue treatment. 
One modest way to deal with this is to communicate to the patients still in line for treatment the current length of the run of failures and to leave it to mutual consent on whether patients want to take the risk of a treatment. 

What would be a good alternative?
Suppose that we had a well-defined inference-based probability measure $$\hat P_\ell(S_n>0):=\hat P(S_n>0|S_\ell=0), ~ \ell \in \{1,2, \cdots, n-1\}.$$ 
Then we could try to fix in agreement with the patients some value $\alpha_\ell$ and rule to stop on the first $\ell \in \N$ with $\hat P_\ell(S_n>0)<\alpha_\ell. $ However, without additional assumptions, any choice of such a probability measure $\hat P_\ell$ seems difficult to justify. Hence, as the author understands, for a beginning run of failures, the mutual agreement should overrule attempts of trying to do better by a questionable quantification.

\subsection{Protocol 2: Unknown success probabilities}
With the unknown odds of successes, and thus the requirement to estimate these sequentially, an adequate algorithm will now have a different structure from the one for Protocol 2. However, to keep it appealing to the physician, we can bring it in a very similar form.  Recall again that
the rule according to inequality (15) now reads: \\
Stop~after completion of the~$k$th~treatment, $~ 1 \le k \le n,$~if\begin{align}\sum_{j=k+1}^n \frac{h_j S_k}{H_k  -h_j S_k} <1,\end{align}
where $S_k$ denotes the number of successes up to step $k$, $H_k:=h_1+h_2+\cdots, h_k$ and $\hat p_k:=S_k/H_k.$ 

\subsection{Algorithm for Protocol 2}
The first two lines of known constants can be written before beginning the first treatment:
$${\rm(i)}~~~~~h_1~~~~~~~~~h_2~~~~~~~~~~h_3~~~~\cdots~~\cdots~~~$$
$$~{\rm(ii)}~~~~H_1~~~~~~~~~H_2~~~~~~~~~H_3~~~~~~\cdots~~\cdots~~~$$
When treatments begin, a third line will collect the current total number of successes  known after the $k$th treatment. This is the essential online ingredient of the algorithm.
$$~~~{\rm(iii)}~~~~~S_1\,~~~~~~~~S_2~~~~~~~~~~S_3~~~~~~~\cdots~~\cdots~~~~$$ The remaining  line (iv) is a deterministic function of the first three, namely
$${\rm(iv)}\,~~~~~\hat r_1~~~~~~~~~~\hat r_2~~~~~~~~~~\hat r_3~~~~~~\cdots~~\cdots~~~$$ which are given by the successive summands in (16). \qed

\bigskip \noindent We need no separate line for the $\hat p_j$'s and $\hat q_j$'s in order to decide when to stop since (see (16)). All what is needed is in the lines (i)-(iii).

\subsection{Protocol 3}
As mentioned before, the decision makers (doctor and patients) may convene that the sequence should be stopped (that is, no further treatments) if a beginning run of failures seems to become too long. For long beginning failure runs we cannot propose a definite decision help, because the beginning $S_j/H_j$'s are then zero so that the $\hat p_j$'s are all zero.

However, as soon as there is at least a success, the $\hat p_j$'s become positive, and then the algorithm can be used to gain additional decision help. For example, the figures may indicate that the chances of at least one more success are still giving sufficient motivation to continue. Therefore we propose:

\subsection {Algorithm for Protocol 3}
Add to the  algorithm for Protocol 2 with (i), (ii), (iii), (iv) the lines
$${\rm(v)}~~~~\hat p_1~~~~~~~~~\hat p_2~~~~~~~~~~~\hat p_3~~~~~\cdots~~\cdots~~~$$
$$~~~~~~~~~~~~~~{\rm(vi)}~~~~\hat q_1~~~~~~~~~\hat q_2~~~~~~~~~~~\hat q_3~~~~~\cdots~~\cdots~~~~~~~~~~~~\qed$$ 
These lines can be filled in sequentially after observing $S_1, S_2,...$.  Note that they can be used to obtain the following information:

\medskip
In line (v): The sum $\hat p_{k+1}+\hat p_{k+2}+\cdots+\hat p_{n}$ represents an estimate of the expected number  $\E(S_n-S_k)$ of further successes after step $k.$

\smallskip
In line (vi): The product $\prod_{j=k+1}^n \hat q_j$ estimates the probability of $P(S_n-S_k=0),$ that is, the probability of no further successes existing after step $k.$

\smallskip
Note that the preceding items of information are of interest in view of the {\it informed consent} requirement usually seen as necessary in any expanded use program for drugs.

\subsection{Optimality} In the case of unknown success probabilities no claim is made that the algorithms yield optimality. We have proposed estimates for the $p_j$'s and $r_j$'s and then applied an online version of the odds-algorithm. Recall that the latter has been proved to be optimal for the {\it true}  $p_j$'s and $r_j$'s, but finding for all $n$ an overall optimal treatment strategy for unknown success probabilities stays an open problem. Now, using the law of large numbers it is easy to see that our solutions through Protocols 2 and 3 are asymptotically optimal as $n$ increases. This is assuring, but $n$ is typically not very large in practice. Nevertheless, simulations in Bruss and Louchard (2009) show that the procedure we propose is good to very good. 
\subsection*{Protocol 4: Unknown stream of requests.}

This Protocol is a  suggestion for a physician (with mathematical inclination)  to establish his or her own protocol in very general situations to obtain  decision help.  Often enough, neither individual success probabilities for patients nor their number are known. Such situations occur for instance if the horizon (real time interval $[0,t]$, say) in which the treatments will be tested had to be fixed but where the physician does not necessarily know how many patients will join the queue up to time $t.$ Patients are thought of as arriving in time according to a stochastic arrival process.

\smallskip
We cannot expect as much decision help as we obtained from the algorithms the Protocols 1. to 3.  since we have now much less information. As we will see below, it is not really the randomness of the number of patients which is the problem for proposing a good model. A more serious drawback is the unknown health scores $h_j$ in the "future"; the doctor cannot assess the health of unknown patients. Hence the $h_j$'s must be replaced by some estimated mean $m_h(u)$ as seen at time $u$ which weakens conclusions. Please note however, that this fact lies in the nature of the problem and should therefore not be considered as a drawback in any method the physician would propose.

\smallskip
\smallskip We suggest here just one way to model the problem, namely, using a Poisson arrival process:

\medskip

$\lambda(u)= $rate of a (inhomogeneous) Poisson processes at time $u\in [0,t]$

\smallskip

${\cal N}_u=$ number of patients arriving up to time $u$

\smallskip
${\cal S}_u=$ number of successfully treated patients until time $u.$

\smallskip
$m_h(u)=$ mean state of health score of patients seen up time $u.$

\smallskip
${\cal P}_u\left({\cal N}_u, {\cal S}_u\right)=$ predicted internal success probability for the treatment of a patient arriving in $]u,t],$ knowing the history of arrivals and successes up to time $u.$ 

\smallskip
The arrival rate or intensity $\lambda(u)$ should be estimated according to the doctor's experience. If no additional information is available, the doctor may see any arrival time as equally likely and model $\lambda(u)$ as being a constant function over the horizon $[0,t]$. For instance, if he thinks that he may expect a total of roughly $r$ requests (arrivals) in $[0,t],$ then he may use $\lambda(u):= r/t.$

The estimates used for ${\cal P}_u$ can the be obtained similarly to our approach in Protocol 2 if we interpret the arrival times of patients as the times when the evaluations of the treatments in success or failure are available. However, in a specific situation the physician may be well in a position to use  estimates based on more information.

\bigskip\noindent
{\bf Suggested strategy for Protocol 4}:

\smallskip
\noindent Refuse new treatment requests from the first time $s$ onwards with\begin{align}\int_s^t
\lambda(u) {\cal P}_s\left({\cal N}_u, {\cal S}_u\right) m_h(s)\, du \le 1. ~~~ \qed\end{align}

\medskip

\noindent{\bf Justification of the suggested strategy}:

\smallskip
\noindent To understand this suggestion we refer to section 4.1 on page 1389 of Bruss (2000) combined with Lemma 5.1 on page 3252 of Bruss and Yor (2012).

 Actually, it is not crucial that the arrival processes is a Poisson process. What is essential is that this arrival process is a process with independent increments, and the Poisson process has this property.  But we propose to use a Poisson process because then the mentioned references Bruss(2000) and Bruss and Yor (2012) directly explain our suggested strategy. Moreover, a Poisson process is particularly convenient model. 
 
 \smallskip
 With independent increments it is of course understood that the intensity $\lambda (u)$ of the Poisson arrival process does not depend on previous successes and failures. Unless $[0,t]$ is large, in the order of two years or more, this assumption can be well defended.
 
  \smallskip
According to a few simulations, the quality of such a strategy as suggested in Protocol 4 seems good if the expected number of arrivals is not too small. In practice the stream of arrivals in compassionate use programmes can be very small, and the author regrets not to be able to make a better suggestion for such a case. For more standard treatments with approved drugs and known side effects, this seems much less important, of course.

Having said this, there are two sides to the coin. Thinking specifically about the context of compassionate use, whether being the physician or a possible patient, it is equally comforting to see cases where only very few patients have to face such serious treatments.

\subsection*{Conclusion} Stopping a sequence of treatments  with the {\it very last success} is, from a medical ethical point of view, an objective which stands out. Stopping on the last success realises all possible successes and minimises the number of future futile treatments. A conscientious physician without prophetical capacities cannot but translate this objective into the objective  of maximising the probability of stopping with the very last success. This paper proposes Protocols to find the {\it optimal} or close to optimal solution in an easily computable and organised way. These protocols also allow for lower safety thresholds to control for
the risk, due to randomness, of continuing too long in waiting for improbable further successes. Finally, the article also discusses in  Protocol 4 a suggestion to deal with the difficult case of an unknown stream of requests for treatments.\subsection*{References}
~~~~Angelos, P.~{~\it The ethical challenges of surgical innovation for patient care}, Lancet, Volume 376 (9746), 1046-1047, (2010)

\smallskip
Ano K., Kakinuma H. and Miyoshi, N. {\it ~Odds Theorem with Multiple Selection Chances},     J. Appl. Probab. ,Vol. 47, Issue 4, pp. 1093-1104 , (2010)

\smallskip
Bruss F. T. {\it~ Sum the Odds to one and Stop},  Annals of  Probab., Vol. 28, No 3, 1384-1391,  (2000).

\smallskip Bruss F. T. {~\it  A Note on Bounds for the Odds-Theorem of Optimal Stopping.}
Annals of  Probab, 
Vol 31, No 4, 1859-1861, (2003))

\smallskip Bruss F. T.  {~\it The Art of a Right Decision: Why decision makers may want to know the odds-algorithm}, Feature art., Newsl. Europ. Math. Society, Issue 62, 14-20, (2006) 

\smallskip Bruss F. T. and Louchard G.~ {\it The Odds-algorithm Based on Sequential Updating and its Performance}, Adv.  Appl. Prob., Vol. 41, 131-153, (2009).

\smallskip Bruss F. T. and Yor M. {\it~ Stochastic Processes with Proportional Increments and the Last Arrival Problem}, Stochast. Proc. \& Th. Applic., Vol. 122, 3239-3261, (2012)

\smallskip  Bunnik E. M., Aarts N. and  van de Vathorst S., {~\it The changing landscape of expanded access to investigational drugs for patients with unmet medical needs: ethical implications}, J. Pharm. Policy Pract., Vol. 10, doi: 1186/s40545-17-0100-3, (2017)

\smallskip Darrow J. J., Sapatwari A., Avorn J. and Kesselheim A.S. {\it ~Practical, Legal, and Ethical Issues in Expanded Access to Investigational Drugs}, New England J. of Med., Vol 372: 279-286, (2015)

\smallskip Dendievel R., {\it New Developments of the Odds-algorithm of Optimal Stopping}, The Math. Scientist, Vol. 38, (2), pp.111-123, (2013).

\smallskip Ferguson T. S.{\it ~The Sum-the-Odds Theorem with application to a stopping game of Sakaguchi}, in Special Volume i.\,h.\,of F. Thomas Bruss, Mathematica Applicanda,Vol. 44(1), 45-61, (2016).

\smallskip  Kesselheim A.S., Robertson C.T.,  Myers J.A., Rose S.L., Gillet V., Ross K.M.,  Glynn R.J., Joffe S. and Avorn J., {\it A Randomized Study of How Physicians Interpret Research Funding Disclosures}, New Engl. J. Med., Vol 376, 1119-1127, (2012)

\smallskip MacKenzie C. R. {~\it Professionalism and Medicine}, HSS Journal, Vol.  Sept. 3 (2), 222-227, (2007)

\smallskip Miller J. E., Ross J. S.,  Moch K. I. and A. Caplan {\it~Characterizing expanded access and compassionate use programs for experimental drugs},
BMC Research Notes, Vol. 10, doi:10.1186/s13104-017-2687-5, (2017)

\smallskip  Sritharan K., Russell G.,  Fritz Z., Wong D.,  Rollin M., Dunning J., Morgan P. and Sheehan C.{~\it Medical oaths and declarations}, British Med.\,Journ., 323 (7327), 1440-1441, (2001)

\smallskip Tamaki M., {\it~Sum the Multiplicative Odds to One and Stop}, J. Appl. Probab.,Vol. 47, Issue 3, 761-777, (2010)

   \centerline{---}
\bigskip\noindent
{\bf Author's adress}: 

\noindent Unversit\'e Libre de Bruxelles, \\Facult\'e des sciences, \\Campus Plaine, CP 210, B-1050 Brussels.\\ (tbruss@ulb.ac.be)
\end{document}